# Molecular Gas Optics Using Vibrational Population Density Refraction

Daniel Matteo, Sergei Tochitsky*, and Chan Joshi

Neptune Laboratory, Department of Electrical Engineering, University of California Los Angeles, Los Angeles, CA, 90095; *Contact author: sergei12@ucla.edu

**Abstract.** Gas optics holds promise for development of advanced optical elements with 2-3 orders of magnitude higher damage threshold than any solid-state optics. Here, we demonstrate a new concept for refracting the laser beam using vibrationally excited molecules as gas optics instead of traditional control of thermodynamic gas temperature. The transient, vibrational population density grating is formed in a $CO_2$ gas by interfering two mid-IR lasers to imprint spatial modulation in the linear refractive index ($\Delta n_0 \geq 2\text{x}10^{-5}$) of the neutral gas via enhancement of the molecular polarizability. It is increased by pumping the asymmetric stretching mode of $CO_2$ at optimal wavelengths to resonantly produce a highly nonequilibrium state that increases the effective vibrational mode temperature to greater than 2000 K in a room temperature gas. The measured beam steering efficiency of such a volume gas grating exceeds 90% that makes this approach a candidate for producing other large-aperture, high-throughput molecular optics elements, with expected damage thresholds ~0.5-1 kJ/cm$^2$, much-needed for future ultrahigh-power lasers.

For centuries, control over the direction, intensity and shape of light was successfully accomplished by tailoring amplitude and phase of light beams with solid-state optical surfaces and bulk materials for which the linear refractive index $n_0$ can be easily manipulated in the range of $10^{-1}$-$10^{-3}$. However, modern photonics is increasingly facing applications where solid-state optics are not applicable anymore, either due to wavelength absorption, laser-induced damage, or optical nonlinearity limitations. In one such application, the next generation of laser driven fusion, particle acceleration, and high-energy physics facilities require large-format high-power optics with a high laser induced damage threshold ≥1 kJ/cm$^2$ (plus power densities >10kW/cm$^2$), and a high uniformity across a 10x10 cm aperture [1]. This remains a significant challenge and calls for development of new concepts for high-power laser steering and focusing optics. At present, there are two media that can potentially withstand ultrahigh-power laser beams: plasmas and neutral gases.

Plasma optics use the polarizability of free electrons to modulate the refractive index. When a high-density plasma is created in the interaction of an intense pulsed laser with a gas or solid surface, the typical value of the refractive index change, $\Delta n_0$, is around $10^{-1} – 10^{-3}$, similar to the values of conventional optical coatings. Plasmas driven by ubiquitous UV or near-IR lasers, have demonstrated diffracting, guiding, and focusing laser beams [2-10] To date, plasma gratings [5-8,10], lenses [9] and mirrors [4] have been proposed for high-power laser systems, with some already being tested. However, this approach is intrinsically unstable due to the nonlinearity of the ionization process and rather energy consuming due to the ionization threshold of 10-15 eV per atom. Plasma optics also have typically a low throughput and a short lifetime [6].

Employment of neutral gaseous media for photonic devices offer several advantages; they are easier to control, possess a high throughput, and have superb spatial homogeneity. The control of refractive index is based on modulating thermodynamic gas temperature, T, and density gradients in a gas medium [11]. To this end, the acousto-optics effect in gases was utilized recently for building an efficient Bragg grating in air [12] and ozone [13]. In the former experiment, applying a conventional ultrasound wave produced a small $\Delta n_0 \sim 10^{-7}$ in air, compensated by a rather long ~0.5 m interaction length to achieve 50% scattering efficiency [12]. In the latter demonstration, two interfering UV laser beams generated a spatially modulated gas heating pattern via laser-controlled photodissociation of ozone, launching acoustic waves in the gas and producing a transient thermal grating [13,14]. For such a volume gas grating, a record 1.6 kJ/cm$^2$ damage threshold was measured for a 0.5 μm probe beam with a scattering efficiency of >90% corresponding to $\Delta n_0 \sim 10^{-4}$-$10^{-5}$. Note, that $O_3$ is a toxic gas that requires electrical discharge in $O_2$ for its generation; both factors hinder the use of this approach in large-scale settings. Thus, gas optics holds promise for building optical elements that have 100-1000 times higher damage threshold than state-of-the-art dielectric coatings but its potential is yet to be fully realized.

We propose a novel approach for realizing gas optics based on control of the vibrational population of molecules characterized by the so-called internal vibrational mode temperature. These processes occur in a nonequilibrium neutral gas prior to the acousto-optic caused thermalization. Using the interference pattern produced by two intersecting MIR lasers resonantly pumping molecules, it is possible to generate a Vibrational Population Density Grating (VPDG) by selective excitation of a vibrational mode. This results in an increase in the linear refractive index of a neutral gas via control of molecular polarizability, $\alpha(\omega)$, since $n_0 = 1 + \frac{\alpha(\omega)}{2\varepsilon_0} N(\omega)$ where $\varepsilon_0$ is free space permittivity and $N(\omega)$ is number of molecules in the resonantly excited state [15]. In general, each resonantly excited vibrational mode can be thought of as an anharmonic oscillator. As explained in Fig.1a, vibrationally excited molecules become larger due to an average bond length increase and therefore possess enhanced polarizability. As a result, molecules in a higher vibration state contribute more to the overall polarizability and hence the refractive index than the molecules in a ground state. This can form the refractive index grating with extremely high diffraction efficiency. Such optical excitation may provide a means to control $n_0$ of more than a dozen simple dipole molecules, such as$CO_2$, CO, $C_2H_4$, and $SF_6$ [15] and enable large-aperture gas optics.

Resonant excitation of molecular vibrations in the range of 2-20 μm wavelengths was studied in the context of intra- and inter-mode molecular kinetics [16,17], for lasing in optically pumped molecular lasers in the MIR/THz range [18,19], nonlinear laser vibrational spectroscopy and microscopy [20-23] and for formation of so-called laser-induced thermal gratings in a gas [24-26]. Previous work has shown that in certain molecules, internal degrees of freedom such as vibration can equilibrate within a mode on time scales which are considerably shorter than those required for the vibrational-translational (V-T) thermalization [27,28]. These phenomena can be described by using a vibrational temperature, $T_v$, which accounts for the relative distribution of vibrational mode energy after laser excitation and can be much higher than the familiar translational temperature of a gas. In excited states populated according to Boltzmann statistics, the refractive index change of a molecular gas has been calculated to scale linearly with $T_v$: $\Delta n_0=\gamma T_v$, where $\gamma$ is a species-dependent temperature coefficient [29]. Experimentally, $\Delta n_0=6\text{x}10^{-8}$ was interferometrically measured in 2 Torr of $SF_6$ excited by a 10 μm laser to $T_v$~1500K [29]. In the case of $CO_2$ gas, high vibrational temperatures have been achieved via excitation of the asymmetric stretching mode,$\nu_3$ to temperatures $T_3 \geq$ 3000K using a narrow band 4.3 μm $CO_2$ laser for pumping [30]. However, to obtain such high-level of mode excitation, this required reduced concentrations of $CO_2$ below ~20 Torr and a long interaction length to compensate for low absorption [30]. This leaves an open question whether vibrational molecular optics is capable to precisely control the refractive index in a gas in the range of $10^{-4}$-$10^{-5}$ needed to produce a high-diffraction efficiency.

In this paper we demonstrate that selective excitation of the asymmetric stretching mode,$\nu_3$ of $CO_2$ molecules with a 4.3 μm laser pulse can enhance the molecular polarizability and increase the linear refractive index of the gas by $\Delta n_0 \geqslant 2\text{x}10^{-5}$. We show that a volume refractive index grating imprinted in neutral $CO_2$ gas by two intersecting pump beams resonant to the $\nu_3$ mode can be extremely efficient in steering a second probe laser beam with the diffraction efficiency ≥92%. Detailed measurements of the vibrational temperature $T_3$ of the $\nu_3$ mode reveal that a critical condition for reaching high diffraction efficiency in such a VPDG is a multi-step and multi-resonant (involving several vibrational states) excitation of the molecular mode when multiple-frequency components of a broadband MIR laser pulse can compensate the frequency red-shift caused by the anharmonicity of vibrations (Fig.1a) and populate multiple levels resulting in high $T_3$ in the maxima of the interference pattern. We find that control of refractivity is related to reaching effective *vibrational* mode temperatures above 2000K in a room temperature gas and the effect persists on the vibrational-vibrational (V-V) exchange scale of ~100s ns indicating significance of intramode kinetics for achieving the peak in $\Delta n_0$.

In Fig. 1b we show a partial energy level diagram of a $CO_2$ molecule including the asymmetric stretching mode, $\nu_3$ with a ro-vibrational spectrum centered around ~4.3 μm. It can be efficiently pumped optically by absorbing up to 9 photons prior to reaching the dissociation level. Due to anharmonicity of this mode, each of 00n→00n+1 (n=1,2,3…) vibrational transitions is slightly red-shifted to longer wavelengths from the 00(n-1) → 00n transition. The large dipole moment of ~0.33 D [31] makes the $CO_2$ $\nu_3$ mode a good candidate for refractivity control studies since many millijoules of pump pulse energy can be absorbed over a few cm-long gas column at atmospheric pressures.

In Fig.1c we present a schematic of the gas optics concept. A Fe:ZnSe laser with spectral tunability in the range of 3.9-5 μm and a full width at half maximum (FWHM) spectral linewidth of ~2 nm [32] provided an experimental platform for optical pumping any part of the $CO_2$ $\nu_3$ spectrum. The ~20 ns FWHM pump laser pulse was split in two and intersected in a cell filled with $CO_2$ gas at an angle of ~6 degrees. Two beams at the central wavelength λ=4.356 μm formed an interference pattern (grating) in a $CO_2$ gas (see inset). Two pump beams with a total energy of 2x15=30 mJ were slightly focused providing ~200 mJ/cm$^2$ peak fluence in the cell. This is orders of magnitude smaller than the plasma formation threshold in $CO_2$ of 0.5-1 kJ/cm$^2$ for nanosecond pulses [33]. A visible diode laser probe beam (λ=658 nm)was scattered off the grating at ~0.8° and analyzed either in the time domain by a photodiode or spatially by a 2D pyroelectric array.

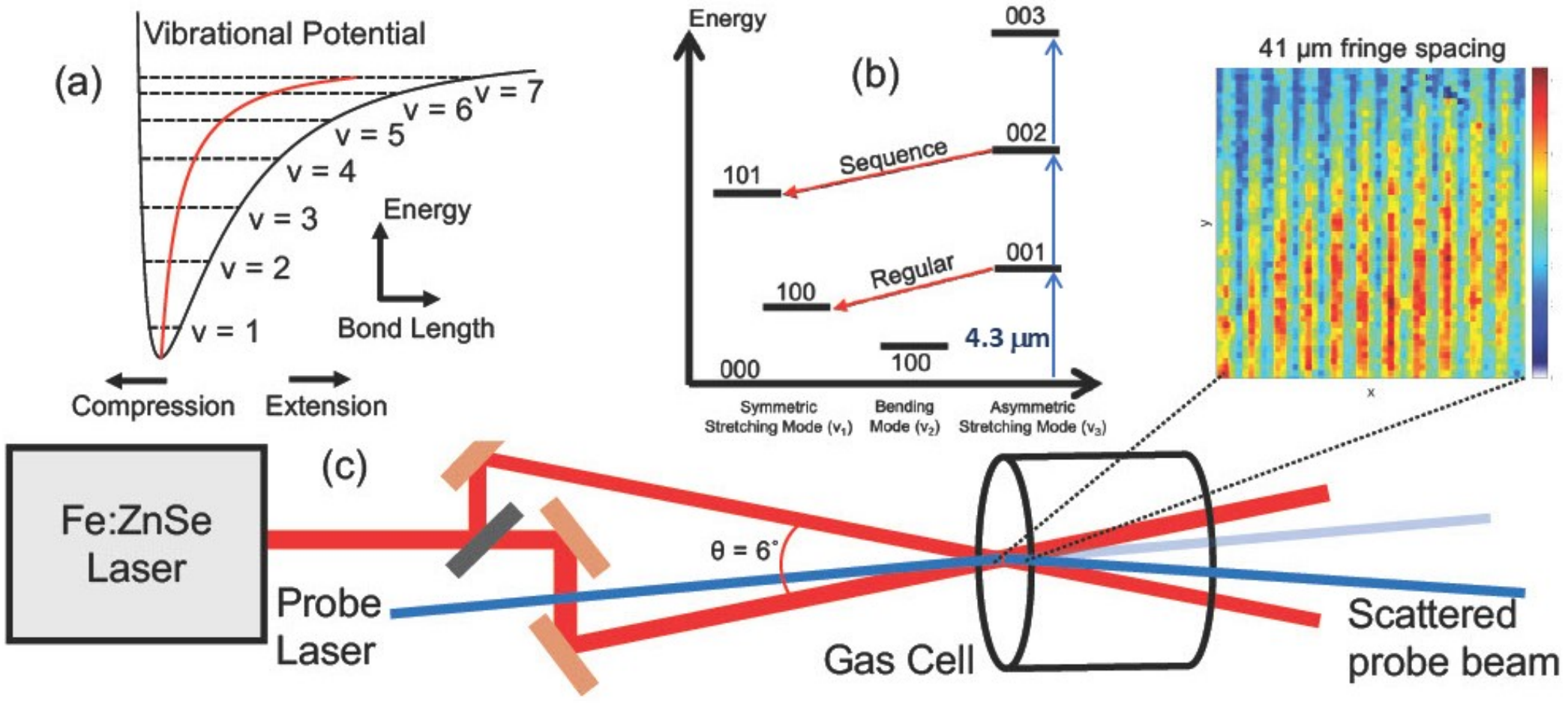

Fig.1 Implementation of optical control of molecular gas refractivity. (a) An example of a molecular potential as an anharmonic oscillator resonantly pumped to high vibrational state. (b) A simplified diagram of $CO_2$ vibrational modes with $\nu_3$ asymmetric stretching mode pumped by a 4.3 μm laser. (c) Experimental set-up. Inset: Interference pattern of two-intersecting pump beams (λ=4.356 μm) in the $CO_2$ filled cell.

The recorded experimental data is presented in Fig.2. Fig.2a shows that the measured scattered probe signal (red curve) in the time domain starts to grow150 ns after the pump pulse (λ=4.356 μm), reaches the maximum value around 400 ns and then drops by a factor of 5 after~1 μs. The fact that the scattering efficiency peaks at ≥100s ns after the pump pulse suggests that intramode vibrational kinetics of the strongly excited $\nu_3$ mode is responsible for the observed scattering because all V-T collisional processes in $CO_2$ molecules are much slower [28]. In the cell filled with 0.5 atm of $CO_2$ (380 Torr) broadened by ~1atm of He, a 170 ns probe pulse interrogates the peak of the scattering and an ≥92% diffraction efficiency in the first order is measured (see Fig.2b). We also observed that while keeping the absorbed energy the same but changing the pump wavelength to 4.2 μm (blue curve), the grating efficiency dropped by an order of magnitude. In this case, the scattering signal increases monotonically in time and saturates after 1.5 μs, thereafter remaining almost constant over further 2 μs. The latter observation indicated that second less efficient and slower mechanism (such as the formation of a classical thermal grating [24]) was at work on the V-T time scale of many microseconds [28]. The measured pressure dependence of the scattering efficiency in Fig.2c clearly peaked around 380±50 Torr of $CO_2$. At pressures of <200 Torr, to preserve the total absorption approximately at the constant level and to study scaling of the probe-light scattering with the number of molecules, we utilized a $^{13}CO_2$ gas. It has a red shifted absorption spectrum of the $\nu_3$ mode [34] and, therefore possessed higher absorption at λ=4.356 μm in comparison with a common $^{12}CO_2$ molecule. Note, that on a high-pressure side above 450 Torr, we began observing multiple orders of diffraction (up to +/-3$^{rd}$ as shown in the inset in Fig.2c) probably related to transition from thick to thin grating regimes [24] with drop in the absorption length.

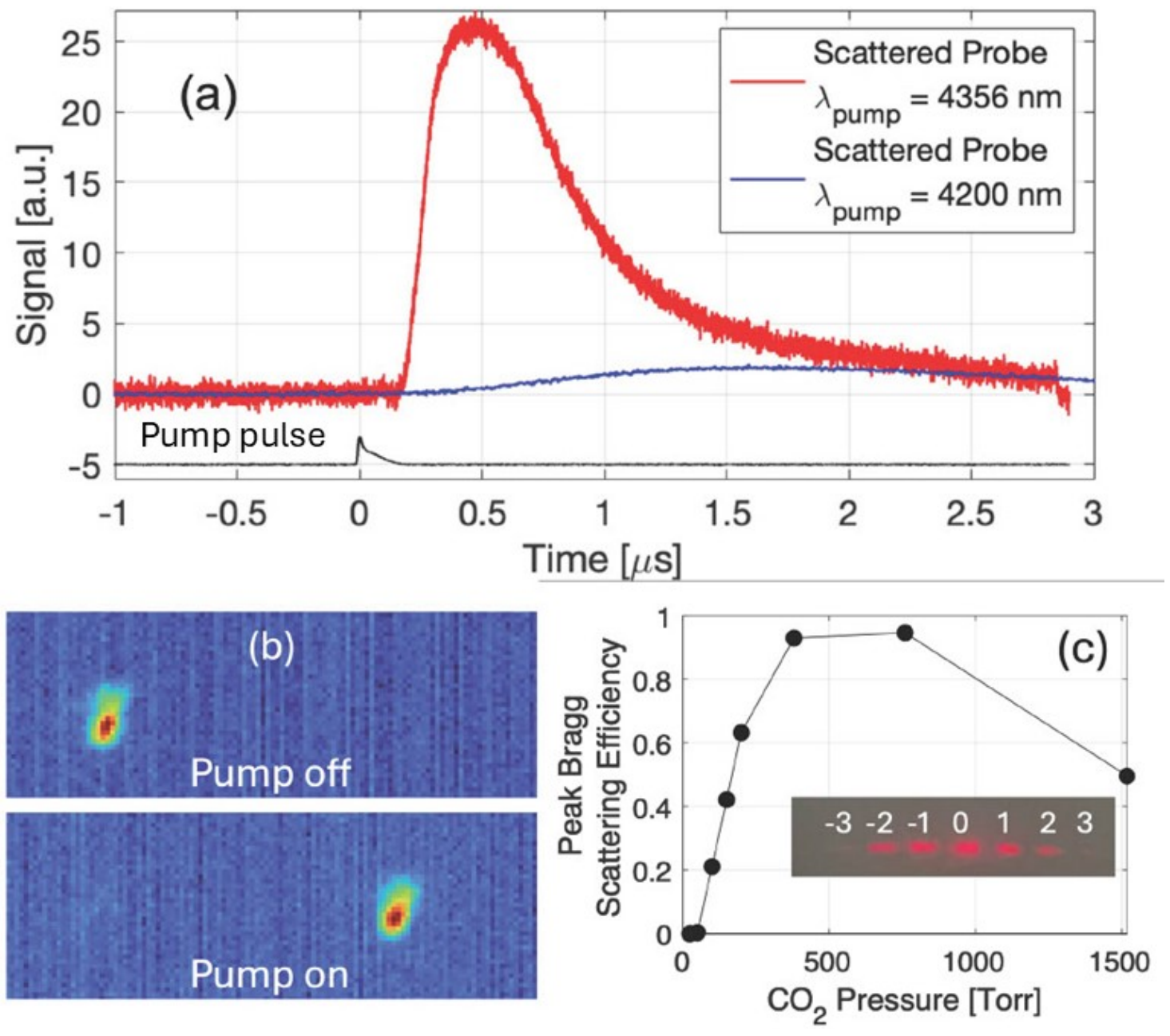

Fig.2 Experimental characterization of a $CO_2$ VPDG. (a) Time domain dynamics as measured by a red diode laser beam. (b) Scattering of the red probe beam in the first order of diffraction at 380 Torr of $CO_2$ broadened by 1 atm of He, (c) Peak diffraction efficiency as a function of $CO_2$ pressure in the cell pumped at λ=4.356 μm. The inset shows scattering in multiple orders recorded for a several μs long probe pulse at 1 atm of $CO_2$.

The observed increase in the probe scattering when pumped on the red side of the $CO_2$ absorption spectrum versus the blue side for the same absorbed pump energy can be explained by investigating the details of the $\nu_3$ mode spectrum. The calculated asymmetric stretching mode (00n) absorption cross-section spectra from [35] are presented in Fig. 3a for the first three vibrational levels. At room temperature, cross-section for the fundamental transition 000–001 is several orders of magnitude stronger than the higher located transitions 001-002 and 002-003, simply due to small 001 and 002 populations at thermal equilibrium. However, as soon as a significant population is created in the 001 level by resonant absorption of photons, the absorption cross-section of the 001-002 transition dramatically increases, enabling strong resonant absorption on the 001-002 frequencies, etc. Additionally, when population is transferred from 000-001 via resonant absorption it gets redistributed across the entire rotational manifold via rotational relaxation during the laser pulse thus avoiding the "rotational bottleneck effect" [28,36,37]. At 1.5 atm, rotational relaxation time is ~100 ps [28]. Therefore, multi-step sequential excitation of $CO_2$ molecules to high lying vibrational levels is possible during a single 20 ns pump pulse, provided the pump laser spectrum overlaps with the higher lying transitions. In Fig. 3b we show how the 2 nm FWHM bandwidth of a pulse centered at 4.356 μm (red line in Fig. 3a) is overlapped with transitions of 000-001,001-002, 002-003 bands, enabling optimal conditions for multi-step (multi-resonant) cascaded excitation of the $\nu_3$ mode of $CO_2$. The situation is completely different for a pump wavelength of 4.2 μm (blue line in Fig.3a); although absorption cross-section can be large, the excitation of higher located molecular levels is hindered by the anharmonicity caused frequency shift. This makes the $\nu_3$ mode excitation by a blue-sided pump inefficient.

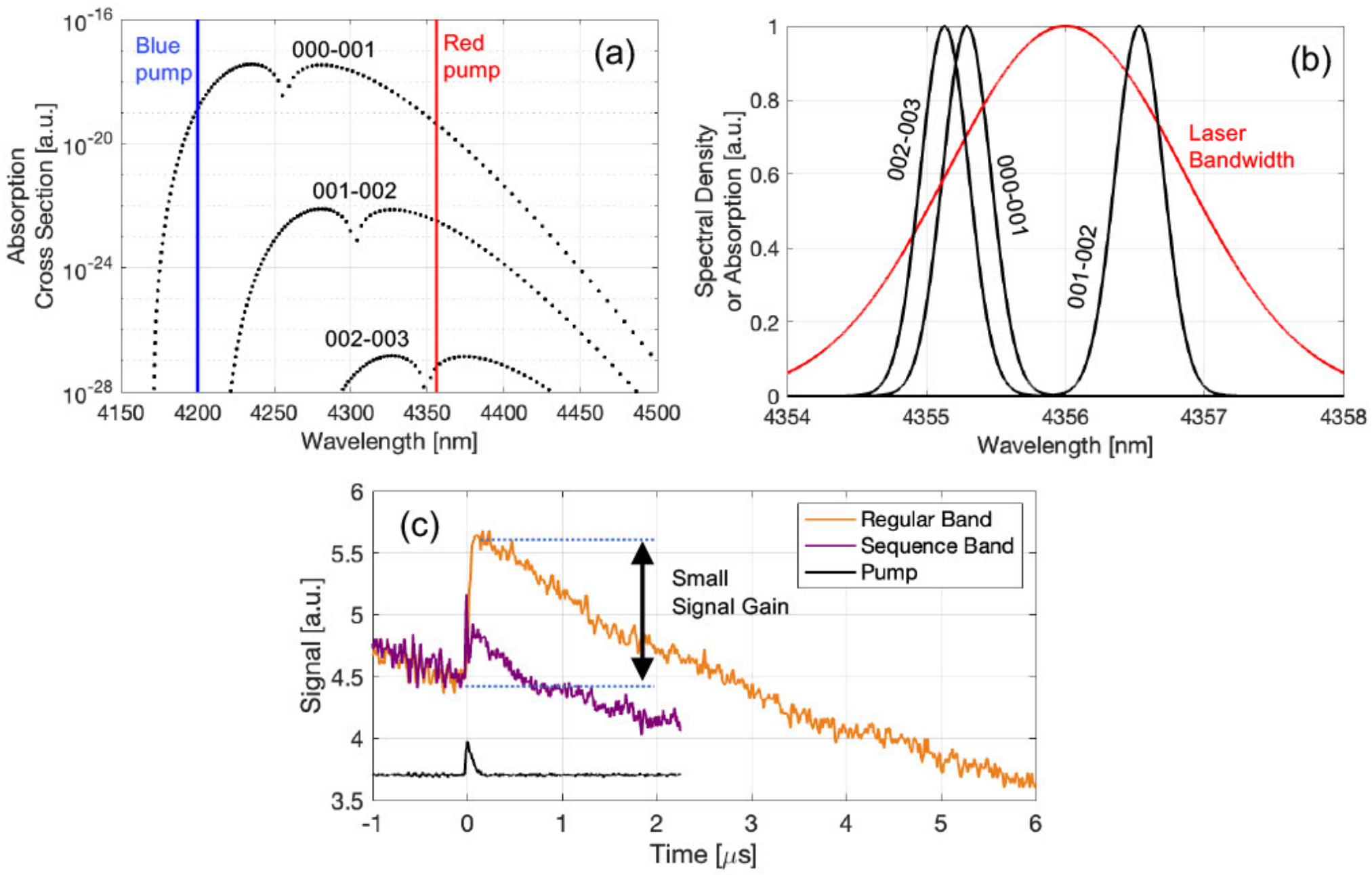

Fig.3 Experimental measurements of vibrational temperatures in $CO_2$. (a) Calculated absorption cross-section for the first three vibrational levels of the $CO_2$ $\nu_3$ mode. Red and blue vertical lines indicate the pump wavelengths used in the experiment. (b) Schematic of the pump laser bandwidth centered at λ=4.356 μm covering several rotational lines of the $\nu_3$ band of $CO_2$. (c) Typical data set for small-signal gain measurements obtained by optical probing of regular 001-100 and sequence 002-101 transitions using a diagnostic 10 μm $CO_2$ laser.

To uncover the physical mechanism behind the observations, we probed 001 and 002 levels of the $CO_2$ molecule using a narrow-line (~200 MHz) diagnostic 10 μm $CO_2$ laser. By tuning between the regular 001-100 and the sequence 002-101 band lines [38] (see Fig.1b), such optical probing allows us to measure the time dependence of the gain and, therefore populations of the vibrational levels directly. Then, assuming a Boltzmann distribution of vibrational population one can calculate the $T_3$ values for the $CO_2$ asymmetric stretching mode using a well-established temperature measurements method developed for $CO_2$ lasers [28,30,38]. In Fig. 3c we present a typical result of small-signal gain measured for the 380 Torr $CO_2$/760 Torr He mixture pumped at λ=4.356 μm. To perform these measurements, we pumped the same gas cell with the same Fe:ZnSe pump laser as above, but with the full energy in a single pump beam. The specific transitions probed were the 10P(26) (10.653 μm) and the 10P(23) (10.647 μm) lines belonging to regular and sequence bands, respectively. The amplified signal was detected by a fast HgCdTe photodetector with ~1 ns resolution. The gain coefficient is then estimated from the peak measured gain, as indicated in Fig. 3c. By measuring the small-signal gain on both regular and sequence bands, the vibrational temperature $T_3$ is calculated by

$$T_3 = \frac{E_3}{k_B}\left(\ln\frac{2g_{reg}}{g_{seq}}\right)^{-1}$$

where $E_3$ is the energy separation between adjacent $\nu_3$ mode levels (~2349 $cm^{-1}$), $k_B$ is the Boltzmann constant and $g_i$ is the measured small-signal gain on each band.

$T_3$ measurements were performed in this way with both the red- and blue-sided pumping of $CO_2$ molecules. Vibrational temperature in the $CO_2$ molecules for the λ=4.356 μm pump was measured to be 2100K approximately 2x more than that pumped at λ=4.2 μm. These results show that when the $CO_2$ medium is pumped on the red side, the pump has enough bandwidth to compensate for the anharmonicity of the $CO_2$ energy structure and create a significant population in all three (001, 002 and 003) excited vibrational states whereas blue side population is mainly created in the 001 level as showed in Fig, 3a. For red-sided pump, in Fig.3c we also observe correlation between the 002-level gain lifetime of ~500 ns with the lifetime of scattering effect in Fig. 2a, again supporting importance of resonant excitation of high vibrational levels. If we take $T_3$=2100K and a literature value for $\chi_{co2}$=1.5x10$^{-5}$K$^{-1}$ [15], then for 380 Torr of $CO_2$ and assuming the interference effect, $\Delta n_0 \approx 2x10^{-5}$. It should be noted that such high $T_3$ values were obtained in a nearly room temperature gas for which translational temperature increased on a microsecond scale from ~300K to 400K [38].

Fig. 4a demonstrates the measured scaling of the gain for the 4.356 μm pumping as a function of the absorbed energy density for 0.5 atm of $CO_2$ in the cell. The nonlinear nature of the gain increase is seen here for the sequence band (purple line). The regular band gain (black line) grows close to linearly with increase in the absorbed energy density, however there is a much faster non-linear increase for the sequence band gain. Therefore, the multi-step excitation can be viewed as a nonlinear excitation effect, where the magnitude of the absorption to higher lying levels depends on the number of photons absorbed in the lower energy transition and population of the 002-level grows faster than linear with the absorbed energy. This level is populated additionally by particles cascaded down from higher located levels 003,004 and beyond consistent with observation in Fig.2a that maximum scattering level is reached 400 ns after the excitation pulse. In Fig.4b we summarize $T_3$ measurements. There is a clear advantage in producing high vibrational temperatures for the case of the multi-step red-sided excitation of the $\nu_3$ mode. Reaching $T_3 \geq 2100$K in the experiment at atmospheric pressures of $CO_2$ is crucial for the efficient beam steering effect. It is attributed to the broadband 4.3 μm pump laser that resonantly excites 002, 003 $CO_2$ levels.

Applying the Boltzmann vibrational population distribution in the nonequilibrium $CO_2$ gas is a simplified way of explaining the data shown in Fig.4. As was reported in [30], for high $T_3$ temperatures above 2000K, use of the Boltzmann distribution of populations instead of a Treanor distribution [39], which takes into account anharmonicity factor of the $\nu_3$ mode explicitly, may affect molecular dynamics. Numerical comparison of differences between these two population distributions is presented in the Supplemental Materials [40]. It indicates that for the Treanor distribution at $T_3$=2100K populations of higher located $CO_2$ levels (n>3) are much larger than those predicted by the Boltzmann distribution. In other words, measuring populations of the 001 and 002 levels in the experiment doesn't allow to study this issue sufficiently. Be as it may, the Boltzmann-based vibrational temperature analysis used to analyze the data is adequate for identifying importance of the multi-step excitation mechanism as well as for calculations of the $\Delta n_0 \approx 2x10^{-5}$ value confirmed independently by the measured Bragg scattering efficiency. The role played by highly excited vibrational levels requires further experimental and theoretical study [41].

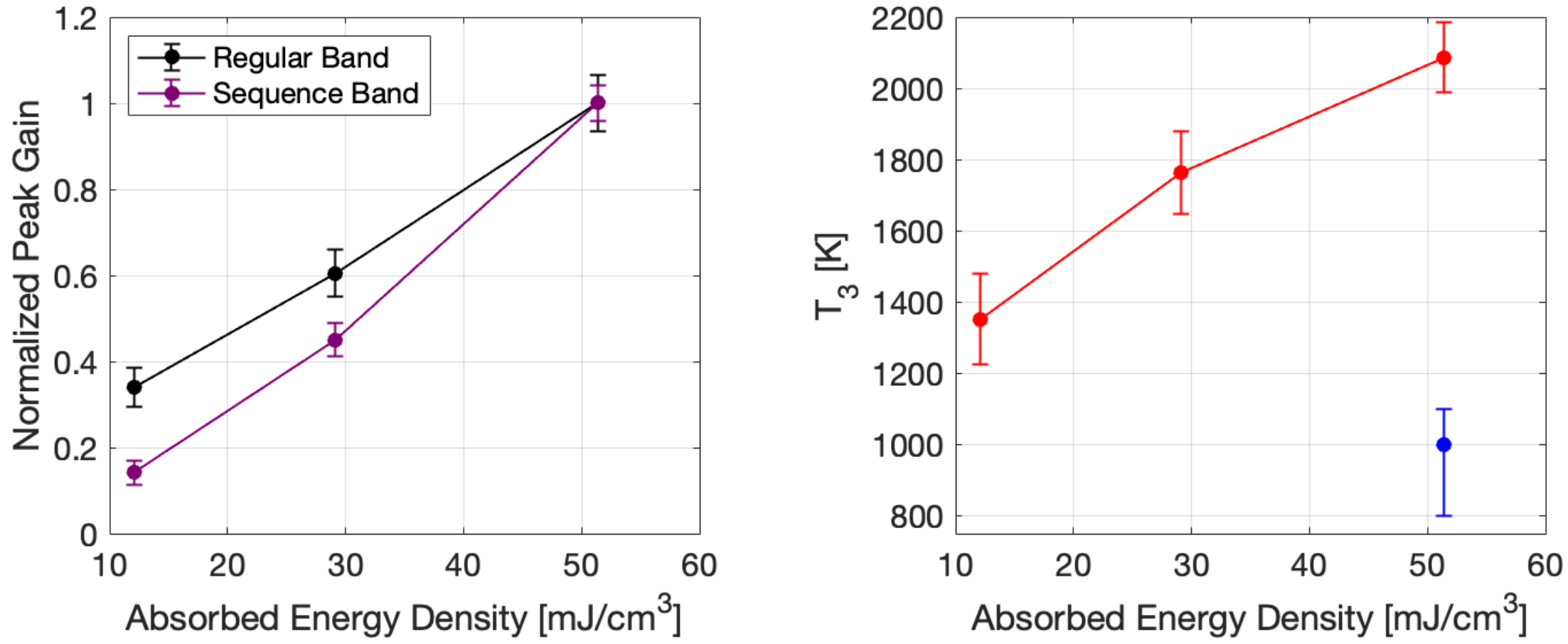

Fig.4 Experimental measurements of vibrational temperatures in $CO_2$ using a line tunable diagnostic $CO_2$ laser. (a) Peak gain measurements for regular 001-100 and sequence 002-101 $CO_2$ transitions as a function of the absorbed energy density at λ=4.356 μm. (b) Vibrational temperature $T_3$ calculated for at λ=4.356 μm (red curve) and at λ=4.20 μm (blue dot) laser pumping in a 380Torr $CO_2$/760Torr He gas mixture.

In conclusion, we report on the first demonstration of vibrational population density grating with the steering efficiency of the probe beam ≥92% achieved in a 4.3 μm-laser excited neutral $CO_2$ gas. Such mechanism of the refractive index control due to high *vibrational* mode temperatures is distinct from the thermal grating in molecules, where the *translational* temperature is modulated on a longer V-T collisional timescale and with a significantly lower scattering efficiency. The described approach for controlling gas refractivity affects all polarizable molecules, suggesting its broad applicability for generation of a new class of MIR laser-controlled large-aperture gas optics isolators and lenses with a ~1kJ/cm$^2$ damage threshold.

**ACKNOWLEDGEMENTS.** The authors are greatly indebted to Prof. S. Mirov, Drs. V. Fedorov and D. Martyshkin from University of Alabama at Birmingham for developing and building a tunable Fe:ZnSe laser. This work was partially supported by DOE grants DE-SC-0018378 and DE-SC-0010064.

**DATA AVAILABILITY**. The data are available upon reasonable request from the authors.

# Supplementary Information: Molecular Gas Optics Using Vibrational Population Density Refraction

Daniel Matteo, Sergei Tochitsky*, and Chan Joshi
Neptune Laboratory, Department of Electrical Engineering, University of California Los Angeles, Los Angeles, CA, 90095

## Details of the Experiment

A Fe:ZnSe laser with spectral tunability in the range of 3.9-5 μm and a full width at half maximum (FWHM) spectral linewidth of ~2 nm [S1] provided an experimental platform for optical pumping any part of the CO2 $\nu_3$ spectrum. The ~20 ns FWHM pump laser pulse was split in two and intersected in a cell filled with CO2 gas at an angle of ~6 degrees. A 10 cm long gas cell was sealed by NaCl windows providing low-loss propagation for 4 μm, 10 μm and red probe beams. Two beams at the central wavelength λ=4.356 μm formed an interference pattern (grating) in a CO2 gas (see inset in Fig.1c). The gas cell was filled with CO2 gas at pressures in the range of 0.01 atm to 1 atm. The ballast He gas was added to keep the total pressure ~ 1.5 atm and mainly served as a collisional partner in order to broaden CO2 molecular lines. This was necessary to partially compensate for absorption of atmospheric CO2 in the transport line from the laser to the cell. Such conditions allowed us to absorb totally approximately ≤20 mJ of the pump light at both 4.356 μm and 4.2 μm wavelengths. In an attempt to increase absorption in the cell by tuning the Fe:ZnSe laser output to stronger CO2 lines, we observed no improvements in the gas grating efficiency, mainly related to detrimental contribution of the atmospheric CO2 absorption. We consider that in the future further optimization and significant decrease in the required pump level (or increasing the aperture of gas grating for the same pump energy) will be possible once the pump transport is organized in N2 gas environment instead of the ambient air. Therefore, total pressure of the CO2-He gas mixture ≥1 atm and gas dynamic windowless arrangement are feasible. The optimal CO2 pressure requirement is very relaxed supporting robust operation of the gas grating. In the experiment, two pump beams with a total energy of 2x15=30 mJ were slightly focused providing ~200 mJ/cm2 peak fluence in the cell. The actual length of the overlapped area for the experimental geometry was 2-3 cm indicating that the cell length can be much shorter.

A visible diode laser probe beam (λ=658 nm) was scattered off the grating at ~0.8° and analyzed either in the time domain by a photodiode or spatially by a 2D pyroelectric array. The measured diffraction angle measured in Fig.2b was very close to the calculated Bragg angle for the first order of diffraction.

## Role of the Distribution Function in Population of Vibrational Levels in a Molecule

For low vibrational temperatures ($T_v \lesssim 1000$ K), molecular vibrational distributions are well-described by Boltzmann statistics, using the typical formulation

$$N_B(\nu) = A_B \exp\left(-\nu \frac{E_\nu}{k_B T}\right)$$

Where N(v) is the population in vibrational mode v, $A_B$ is a normalization constant, $E_v$ is the excitation energy of the vibrational mode, T is the translational temperature, and $k_B$ is the typical Boltzmann constant.

At high vibrational temperatures such as those measured in this work, significant population exists in vibrational levels above the ground state and the red-shifted separation between vibrational levels must be accounted for. This can be done using the Treanor distribution [S2]

$$N_T(\nu) = A_T \exp\left(-\nu \frac{E_\nu}{k_B T_\nu} - \nu(\nu-1)\frac{\Delta E_\nu}{k_B T}\right)$$

Where $T_v$ is the vibrational temperature and $\Delta E_v$ is the energy anharmonicity for the vibrational mode [S3].

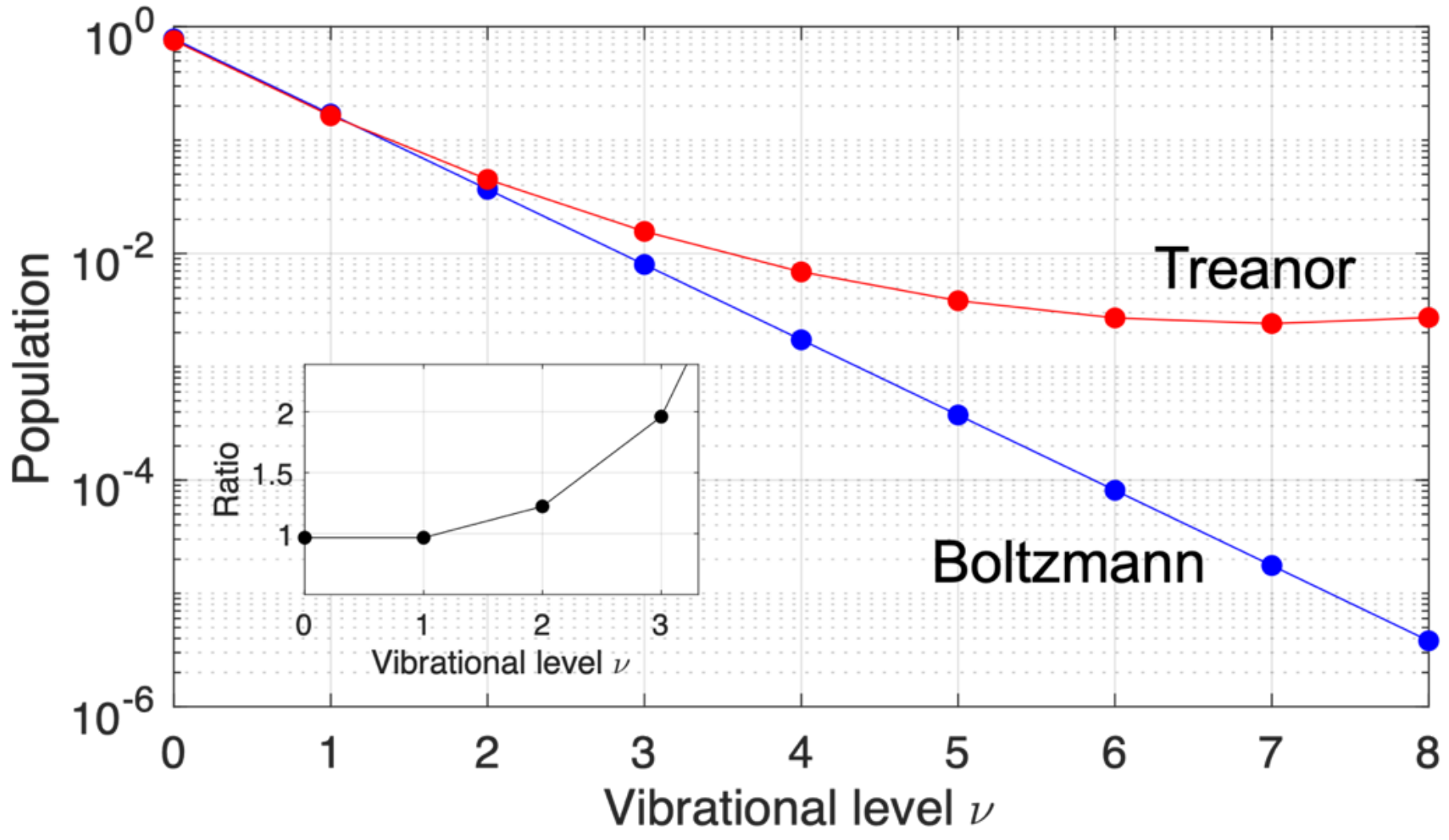

Fig.S1 Calculated population distribution of the CO2 $v_3$ mode using Boltzmann and Treanor distributions at T3=2100K.

Fig. S1 plots both Boltzmann and Treanor distributions in the CO2 asymmetric stretching mode (00v) for the peak vibrational temperature measured in experiment, $T_v$ = 2100 K. Translational temperature is assumed to be room temperature, T = 293 K. At high vibrational levels (v≥3), the Treanor distribution has significantly higher occupation than the Boltzmann distribution, including more than an order of magnitude for v≥5. The figure inset shows the population ratio $N_T/N_B$ for lower lying levels. Since the higher lying vibrational levels are expected to provide larger contributions to molecular polarizability, using the proper Treanor distribution is necessary for accurate analytical correlation of the vibrational temperature and the refractive index change.